# Overview of EIREX 2011: Crowdsourcing

Julián Urbano, Diego Martín, Mónica Marrero and Jorge Morato

University Carlos III of Madrid
Department of Computer Science
Avda. Universidad, 30
Leganés, Madrid, Spain

## 1. Introduction

The second Information Retrieval Education through EXperimentation track (EIREX 2011) was run at the University Carlos III of Madrid, during the 2011 spring semester.

EIREX 2011 is the second in a series of experiments designed to foster new Information Retrieval (IR) education methodologies and resources, with the specific goal of teaching undergraduate IR courses from an experimental perspective. For an introduction to the motivation behind the EIREX experiments, see the first sections of [Urbano et al., 2011a]. For information on other editions of EIREX and related data, see the website at http://ir.kr.inf.uc3m.es/eirex/.

The EIREX series have the following goals:

- To help students get a view of the Information Retrieval process as they would find it in a real-world scenario, either industrial or academic.
- To make students realize the importance of laboratory experiments in Computer Science and have them initiated in their execution and analysis.
- To create a public repository of resources to teach Information Retrieval courses.
- To seek the collaboration and active participation of other Universities in this endeavor.

This overview paper summarizes the results of the EIREX 2011 track, focusing on the creation of the test collection and the analysis to assess its reliability. Next section provides a brief overview of our course and the student systems. Section 3 describes the process we followed to create the EIREX 2011 test collection, and Section 4 presents the evaluation results. Section 5 analyzes the reliability of our approach by studying the effect of the incompleteness of judgments. Section 6 wraps up with the conclusions.

## 2. Teaching Methodology

EIREX 2011 took place during the 2011 spring semester, in the context of the Information Retrieval and Access course [Urbano et al., 2010b], which is an elective course taken by senior Computer Science undergraduates. In this course we teach traditional IR techniques, and the main lab assignment consists in the development, from scratch, of a search engine for HTML documents. The development of this search engine is divided in three modules to hand in separately:

- Module 1 contains the implementation of an indexer for a collection of HTML documents and a simple retrieval model for automatic ad hoc queries.
- Module 2 incorporates query expansion in the retrieval process.
- Module 3 adds simple Named Entity Recognition (NER) capabilities to aid in the "who" questions.

In this edition we had 19 students, who created a total of 5 systems in groups of 4 students each. Thus, we had a total of 15 systems: 5 with basic retrieval, 5 with query expansion and 5 with NER. We try to encourage students by giving one extra point to the group who developed the most effective search engine (see Section 4).

All these systems are evaluated with an IR test collection built with the students also from scratch (see Section 3). A test collection for Information Retrieval evaluation contains three major components: a document collection, a set of information needs (usually called topics), and the relevance judgments or ground truth (usually assessed by humans) telling what documents are relevant to the topics [Voorhees, 2002]. The students run their systems for each topic, returning the list of documents in the collection deemed relevant to it. Then, we use some effectiveness measures to assess, according to the relevance judgments, how well the systems actually answered the information needs. This provides us with a ranking of the student systems in terms of effectiveness.

Last year, we did not have a full training test collection for students to tune their systems, so we provided them with a subset of the EIREX 2010 test collection [Urbano et al., 2011b]. This year, we provided students with the full EIREX 2010 test collection to develop and train their systems. We then evaluated them with the EIREX 2011 test collection and published the results. The same process was repeated for the three modules of each student system.

## 3. Test Collection

The process we employ to create the EIREX test collections is different from those usually followed in other IR evaluation workshops such as the early ad hoc tracks of the Text REtrieval Conference (TREC) ran by NIST [Voorhees et al., 2005]. Although we follow very similar principles, working with students bears some limitations, mainly in terms of effort and reliability. The document collection cannot be as large as those usually employed in TREC, because undergraduate students do not have the adequate expertise to handle that much information and they would probably dedicate too much time to efficiency issues rather than effectiveness and the implementation and understanding of the IR techniques we explain. This limitation restricts the topics to use: if they had nothing in common, we would probably need too many documents to have sufficient diversity to include relevant material for each topic; but if they were somehow similar, probably fewer documents would be needed. Therefore, we decided that all topics should have a common theme, which in addition reflects more closely a real setting where students have to deploy an IR system for a company in a particular domain. Thus, the document collection depends on the topics and not the other way around as usual. For this second EIREX edition we chose the theme to be *Crowdsourcing*, as it is a topic of interest for Information Retrieval.

| Topic | Downloaded | Pool size | Pool depth |
|-------|-----------|-----------|------------|
| 001   | 599       | 101       | 41         |
| 002   | 707       | 100       | 32         |
| 003*  | 486       | 101       | 27         |
| 004*  | 861       | 105       | 19         |
| 005   | 940       | 102       | 19         |
| 006   | 732       | 101       | 37         |
| 007   | 490       | 104       | 13         |
| 008   | 390       | 100       | 32         |
| 009   | 752       | 102       | 25         |
| 010   | 772       | 101       | 20         |
| 011   | 846       | 104       | 20         |
| 012*  | 446       | 100       | 45         |
| 013   | 268       | 100       | 31         |
| 014   | 189       | 101       | 25         |
| 015   | 485       | 102       | 17         |
| 016   | 372       | 101       | 33         |
| 017   | 255       | 100       | 31         |
| 018   | 366       | 101       | 46         |
| 019   | 250       | 100       | 22         |
| 020   | 250       | 102       | 31         |
| 021   | 221       | 100       | 27         |
| 022   | 392       | 101       | 45         |
| 023*  | 647       | 101       | 24         |
| 024†  | 760       | -         | -          |
| 025†  | 769       | -         | -          |
| Average | 530     | 101       | 29         |
| Total | 13,245    | 2,088     | -          |

Table 1. Summary of the EIREX 2011 test collection. * for topics judged by faculty, † for noise topics.

The problem at this point is how to build a document collection making sure that some relevant material is included for every topic. In EIREX 2010 the topics were chosen by the course instructors alone, but this year we involved students in the topic creation too. Each student had to come up with three candidate topics about crowdsourcing. For each topic, they had to issue queries to Google Web Search just as if they were trying to satisfy the information needs themselves, manually using term proximity operators, query expansion, etc.; and report an estimation of the amount of relevant material found in the first page of results. Based on this information we discarded topics apparently too difficult, with very few documents, or for which there did not seem to be clearly relevant documents. Once the final topic set was established, we used a focused web crawler to download all web pages returned by Google for each topic [Urbano et al., 2010a]. The union of all these web pages conform the *complete* document collection.

At this point we have a document collection and a set of topics, so next we need relevance judgments. Another difference here is that students have to make all relevance judgments *before* they start developing, as otherwise some might try cheating and judge all documents retrieved by their system as relevant. In addition, having them inspect the documents to assess their relevance helps later on during development because they know what kind of documents their systems will have to handle. Judging every document for every topic is completely impractical because it requires too much effort, so instead a sample of documents is judged for each topic (i.e. the topics' pools). To come up with a reliable pool of documents despite student systems do not directly contribute, we use well-known and freely available IR tools instead: Lemur[1] and Lucene[2] (call these the *pooling* systems). We thus proceed to index the complete document collection and obtain the results provided by different configurations of the pooling systems for each of the topics, trying to exploit as much as possible our previous knowledge about the topic and the information documents must contain to be considered relevant. For instance, if the topic asked for information about the CEO of a company, we would include the name of the person in the query. With these results we come up with the pools of documents students have to judge.

Pools are formed differently too: if we calculate *depth-k* pools (joining the top $k$ results from the pooling systems), some topics might have considerably more documents to judge than others, as the final number depends on the overlap among the results. If some students were assigned a pool significantly larger than others, they could just judge carelessly once they think they have done enough work compared to their classmates. To prevent this situation we compute *size-k* pools instead: pools with the minimum depth such that the total size is at least $k$ documents. Thus, each topic has a pool of documents with different depth, although all pools have very similar sizes and so all students judge more or less the same amount of documents.

Although unlikely, the results provided by the pooling systems might still leave out relevant documents. To assure that all pools have some relevant material, we always include Google's top $k_G$ results for each topic, as we checked when selecting topics that some relevant web pages were included there. Also, we add $k_N$ random documents crawled from noise topics, which we created by excluding specific terms appearing in the topic set descriptions. These noise documents allow us to check for quality in the relevance judgments, as they should all be judged not relevant for any topic. If we found students judging these noise documents as relevant, we would have an indication of possible negligence. Therefore, all pools have $k_N$ noise documents, the first $k_G$ documents retrieved by Google, and documents retrieved by the pooling systems up to a minimum of $k$ documents altogether. The union of all documents in these pools conform the *biased* document collection. This is the collection we provide students with to run and evaluate their systems.

|  |  | Topics | | Complete Collection | | | Biased Collection | | |
|---|---|---|---|---|---|---|---|---|---|
| Year | Theme | # | Avg. words | # | Avg. words | Size | # | Avg. words | Size |
| 2010 | Computing | 20 | 9 | 9,769 | 1,307 | 735 MB | 1,967 | 1,319 | 161 MB |
| 2011 | Crowdsourcing | 23 | 6 | 13,245 | 1,149 | 952 MB | 2,088 | 902 | 96 MB |

Table 2. Summary of the EIREX test collections.

Table 2 summarizes the size of the EIREX 2010 and 2011 collections. It can be seen that this year we had a larger collection, although the biased version ended up having a similar number of documents but a much smaller overall size. In general, documents were smaller this year, and the topics were more similar (e.g. 52% of the topic titles contained the term *crowdsourcing*). As explained in Section 3.3, this made the pooling systems retrieved documents for a topic that were not retrieved by Google for that particular topic. That is, there was a higher level of overlap between topics, and the biased collection was thus further reduced.

---

[1] http://www.lemurproject.org
[2] http://lucene.apache.org

## 3.1. Topics

The EIREX 2011 test collection contains a total of 23 topics, all of which pertain to the *Crowdsourcing* theme we chose. All topic descriptions have a common structure (see Figure 1) with a unique id, a title and a description of what is considered to be relevant to the topic. Last year we kept things simple and had a generic description of relevance levels for all topics [Urbano et al., 2011b], but this year we made topic-specific descriptions.

```
<topic id="2011-014">
  <title>Use of gold units in CrowdFlower</title>
  <relevance>
    <level value="2">the document must explain what gold units are in Crowdflower and give tips on how they are
        used for quality control.</level>
    <level value="1">the document may explain what gold units are, but it does not mention how they are used or
        should be used.</level>
    <level value="0">the document may mention gold units but it does not explain what they are.</level>
  </relevance>
</topic>
```

Figure 1. A sample EIREX 2011 topic description.

The topic titles were used as input queries to the student systems, so they can all be considered *short automatic* runs in TREC's terminology [Voorhees et al., 2005] (i.e. there is no human intervention in creating the queries from the topic descriptions). Topic titles were significantly shorter than last year (see Table 2): only 6 words on average, ranging from 3 to 11. Topics 024 and 025 were used as noise topics to obtain nonrelevant documents[3].

## 3.2. Documents

The *complete* document collection contains all documents returned by Google for the final set of 23 topics plus the 2 noise topics (see Table 1). A total of 13,245 web pages were crawled for all 25 topics, which account for 952 MB. The median size per document is 47 KB, with a mean of 74 KB. The median number of words per document is 1,149, with a mean of 2,680. These documents were used just as downloaded, with no postprocessing involved. The *biased* collection, containing only documents in the pools (see Section 3.3), had a total of 2,088 documents, which account for 96 MB. The median size per document is 37 KB, with a mean of 47 KB. The median number of words per document is 902, with a mean of 1,435. Compared to last year's, the complete collection has significantly more documents, while the size of the biased collection is significantly reduced despite containing a similar number of documents (see Table 2).

## 3.3. Pools

The document pools were created as in the EIREX 2010 collection [Urbano et al., 2011b]. For each of the 23 topics in the collection, we ran the 12 pooling systems described in Table 3. We used various configurations of Lemur version 4.11 and Lucene version 3.0.1, which basically differed on the stemmer, the treatment of stop words, the retrieval model employed and the use of query expansion.

| Id | System | Parse HTML | Stemmer | Stop words | Model | Query expansion |
|---|---|---|---|---|---|---|
| p0001 | Lemur 4.11 | Yes | Krovetz | No | Okapi BM25 | No |
| p0002 | Lemur 4.11 | Yes | Krovetz | No | Okapi BM25 | Yes |
| p0003 | Lemur 4.11 | Yes | Krovetz | Yes | Okapi BM25 | No |
| p0004 | Lemur 4.11 | Yes | Krovetz | Yes | Okapi BM25 | Yes |
| p0005 | Lemur 4.11 | Yes | No | No | Okapi BM25 | No |
| p0006 | Lemur 4.11 | Yes | No | No | Okapi BM25 | Yes |
| p0007 | Lemur 4.11 | Yes | No | Yes | Okapi BM25 | No |
| p0008 | Lemur 4.11 | Yes | No | Yes | Okapi BM25 | Yes |
| p0009 | Lucene 3.0.1 | No | No | Yes | Vectorial TF/IDF | No |
| p0010 | Lucene 3.0.1 | Yes | No | Yes | Vectorial TF/IDF | Yes |
| p0011 | Lucene 3.0.1 | Yes | Porter | Yes | Vectorial TF/IDF | No |
| p0012 | Lucene 3.0.1 | Yes | Porter | Yes | Vectorial TF/IDF | Yes |

Table 3. Summary of the EIREX 2011 pooling systems.

---

[3] Noise topics have no description in the topics file.

For each topic, we joined the top $k_G$ documents retrieved by Google and $k_N$ random documents from the two noise topics. As last year, we chose $k_G=k_N=10$ documents. Then, we pooled results from the 12 pooling systems until at least 100 documents were in the pool altogether. As shown in Table 1, pool sizes ranged between 100 and 105, with an average of 101 documents. Therefore, all students judged more or less the same amount of documents. Pool depths ranged between 13 and 46, with an average 29, showing that the pooling systems tended to agree much more for some topics than for others. Note that the sum of all pool sizes is 2,330, while the biased collection contains 2,088 unique documents (90%). This indicates that several documents were retrieved for more than one topic: 165 were retrieved for 2 topics, 37 were retrieved for 3 topics and 1 was retrieved for 4 topics. Last year, the biased collection contained 97% of the total maximum documents, so this year we achieved a higher level of overlap across topics.

### 3.4. Relevance Judgments

We applied a cleaning process to all web pages before being displayed to the assessors, turning them into a basic black and white document to make the reading task easier. We also removed all scripts, embedded objects and HTML elements not related to page rendering. Assessors were able to use a basic search option, and they of course did not know whether documents were from the Google top results or noise topics. Judging took a little below 2 hours per assessor and topic, so the task could be completed in one class session. Students never had access to the relevance scores, as all files were encrypted for submission back to the course instructors.

Last year we had two students judge every topic-document in order to measure the effect of inconsistency on relevance judgments, that is, de degree to which using one or another assessor affects the evaluation results [Urbano et al., 2011b][Voorhees, 2000]. We found that students agreed with each other to the same degree TREC assessors do, and that differences between students' judgments did not have a significant impact on the results. Therefore, for this EIREX 2011 edition we had only one student judge the ~100 documents per topic, except for topics 003, 004, 012 and 023, which were judged by one faculty member. We used a 3-point relevance scale from 0 to 2: nonrelevant, somewhat relevant, and highly relevant. Documents that could not be judged due to technical problems when rendering were judged as -1 (0.6% of the times). On average, students judged 20 documents per topic as somewhat relevant and 30 as highly relevant.

Of all the 266 judgments on noise documents, 18 times (7%) was the document judged somewhat relevant and 23 times (9%) was it judged highly relevant. We manually inspected these documents and, for the most part, we found that they were indeed relevant. Therefore, this year's noise topics did not seem to work as well as expected to automatically detect low quality judgments. On the other hand, the fact that we agreed with these judgments indicates that the overall student judgments should be fine too.

### 4. Evaluation Results

All 23 topics were used to evaluate the 15 student systems (3 modules for each of the 5 student groups). We used NDCG@100 (Normalized Discounted Cumulated Gain at 100 documents retrieved) as the main measure to rank systems, and AP@100 (Average Precision), P@10 (Precision) and RR (Reciprocal Rank) as secondary measures, using a 2-point relevance scale conflating the somewhat and highly relevant levels.

Table 4 and the plot in Figure 2 show the mean scores for each of the four measures over the 23 topics. System 04.2 obtained the best NDCG@100 score, 0.486, and therefore received the extra point this year. Systems by groups 04 and 05 rank at the top of the ranking for NDCG@100 and AP@100 (which produce the same ranking). However, group 05 is clearly better to retrieve relevant documents toward the top of the results, as can be observed with the RR and P@10 scores. In fact, system 01.1 is also shown to be better than all systems by group 04 in terms of RR. For all measures, systems by group 02 perform significantly worse than the rest.

Though not as much as last year [Urbano et al., 2011b], student systems again behaved fairly well compared to usual TREC ad hoc results [Voorhees et al., 2005], probably due to the methodology followed to build the test collection (see Section 3). Documents were crawled for a prefixed set of topics, and if topics were quite different from one another (which we attempted to avoid), documents would probably be very different too. That is, it might be somewhat clear, from an algorithmic perspective, what documents pertain to what topics, although systems would still have to rank relevant documents properly. This can be observed in Figure 3. The left plot shows, for each group's best system, the average ratio of documents retrieved at different cutoffs that were crawled for each topic. Call this measure C@k (Crawl). We can see that for most systems about 50% of the documents retrieved were actually crawled for the particular topics. The right plot displays the R@k scores (Recall), showing that the top systems retrieved about 60% of the relevant documents by the end cutoff $k=100$.

| System | NDCG@100 | AP@100 | P@10 | RR |
|---|---|---|---|---|
| 04.2 | 0.486 ± 0.244 | 0.305 ± 0.247 | 0.37 ± 0.277 | 0.503 ± 0.337 |
| 04.1 | 0.48 ± 0.238 | 0.294 ± 0.242 | 0.37 ± 0.275 | 0.5 ± 0.34 |
| 05.1 | 0.478 ± 0.247 | 0.293 ± 0.249 | 0.435 ± 0.316 | 0.551 ± 0.365 |
| 04.3 | 0.469 ± 0.242 | 0.289 ± 0.235 | 0.374 ± 0.263 | 0.531 ± 0.351 |
| 05.2 | 0.462 ± 0.245 | 0.276 ± 0.237 | 0.439 ± 0.323 | 0.567 ± 0.382 |
| 05.3 | 0.462 ± 0.245 | 0.276 ± 0.237 | 0.439 ± 0.323 | 0.567 ± 0.382 |
| 01.1 | 0.449 ± 0.214 | 0.245 ± 0.22 | 0.357 ± 0.25 | 0.547 ± 0.367 |
| 03.1 | 0.422 ± 0.221 | 0.239 ± 0.195 | 0.378 ± 0.266 | 0.455 ± 0.357 |
| 03.2 | 0.422 ± 0.221 | 0.239 ± 0.195 | 0.378 ± 0.266 | 0.455 ± 0.357 |
| 03.3 | 0.422 ± 0.221 | 0.239 ± 0.195 | 0.378 ± 0.266 | 0.455 ± 0.357 |
| 01.2 | 0.386 ± 0.18 | 0.178 ± 0.151 | 0.283 ± 0.219 | 0.492 ± 0.37 |
| 01.3 | 0.369 ± 0.184 | 0.167 ± 0.151 | 0.265 ± 0.217 | 0.475 ± 0.378 |
| 02.1 | 0.235 ± 0.175 | 0.087 ± 0.124 | 0.17 ± 0.172 | 0.329 ± 0.334 |
| 02.2 | 0.223 ± 0.164 | 0.08 ± 0.113 | 0.148 ± 0.156 | 0.316 ± 0.313 |
| 02.3 | 0.096 ± 0.106 | 0.022 ± 0.038 | 0.083 ± 0.111 | 0.189 ± 0.287 |

Table 4. Mean and standard deviation of the NDCG@100, AP@100, P@10 and RR scores for the 15 student systems over 23 topics. Systems are ordered by mean NDCG@100 score.

Last year, most of the best systems obtained C@k scores between 75% and 100% and R@100 scores round 90%. That is, last year it seemed much easier to find the relevant material for each topic. This difference could be attributable to the fact that this year we had a larger overlap among topics (see Section 3.3). However, this year's systems obtained worse results than most of last year's systems with the EIREX 2010 collection, so we believe that; in general, we just had worse systems this time.

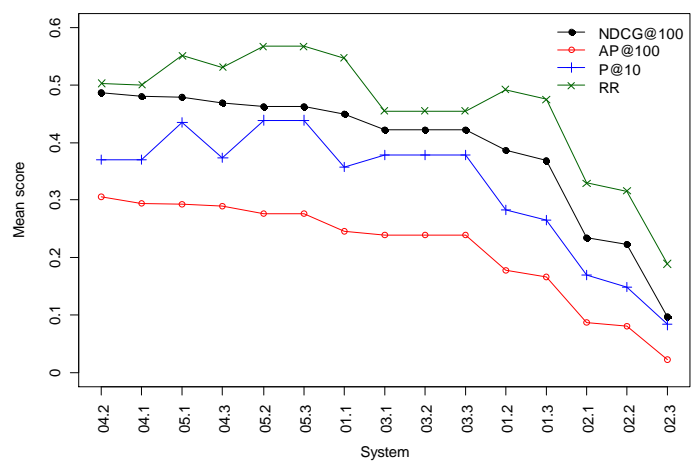

Figure 2. Mean NDCG@100, AP@100, P@10 and RR scores for the 15 student systems over the 23 topics. Systems are sorted by mean NDCG@100 score.

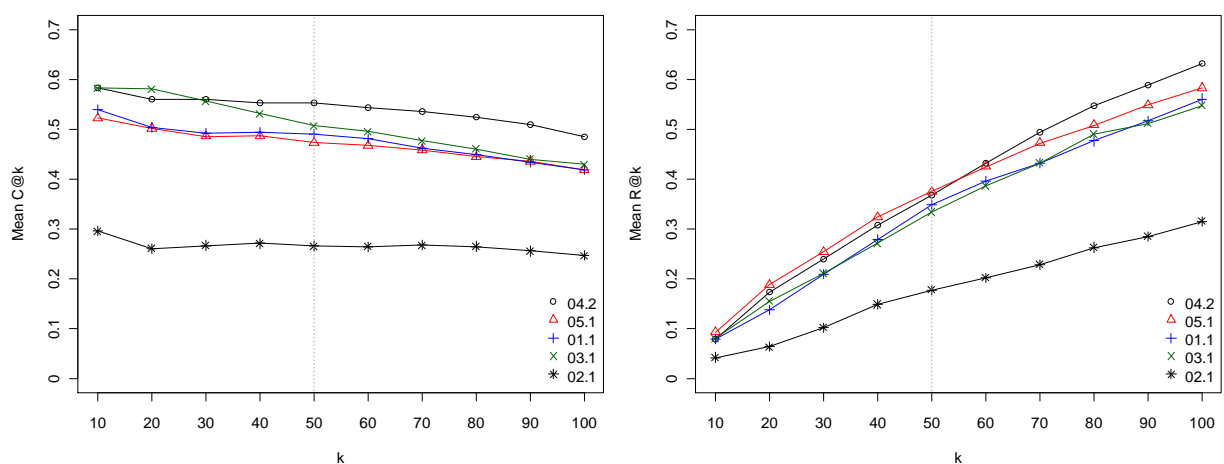

Figure 3. Mean C@k (left) and R@k (right) for the best system per student group. The grey vertical line marks the mean number of relevant documents across topics (50).

# 5. Incompleteness of Relevance Judgments

A drawback of TREC-like evaluations is that the sets of relevance judgments are incomplete, because only the documents in the pools are judged [Voorhees, 2002]. Of course, if a system does not have the opportunity to contribute much to the pool it is expected to have its effectiveness diminished, as it might have retrieved relevant material which is unknown. In the worst case, a brand new system using these collections did not contribute at all to the pools, and so its evaluation could be unreliable. This is our case, as the systems developed by the students did not contribute to the pools, only the Lemur and Lucene systems did (the pooling systems). The effect of incompleteness has been studied with TREC dada, concluding that the early ad hoc tracks were quite robust to the incompleteness problem [Zobel, 1998]. Last year we conducted a similar analysis with the EIREX 2010 collection, and we also found that the small pools we used were quite reliable despite their depth differences [Urbano et al., 2011b]. Next, we analyze the EIREX 2011 collection in the same line.

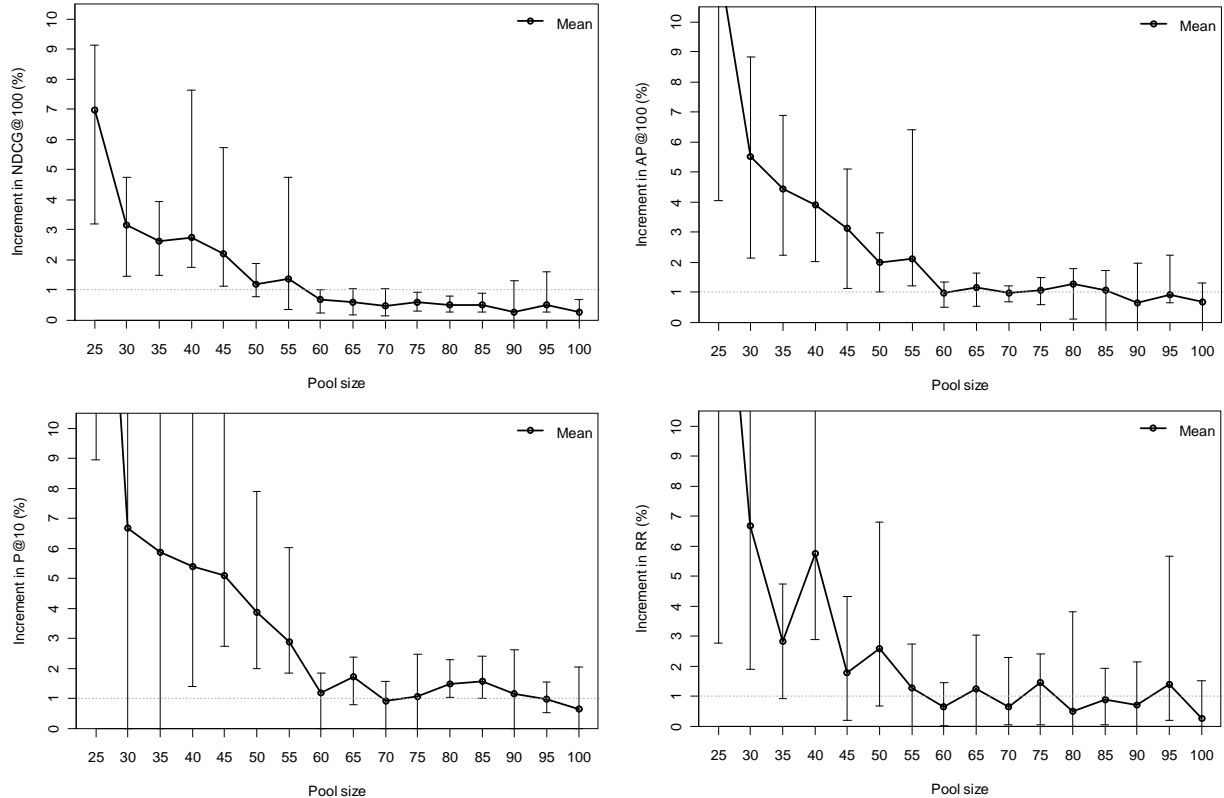

Figure 4. Mean NDCG@100 (top left), AP@100 (top right), P@10 (bottom left) and RR (bottom right) increments as a function of pool size (lower is better). Error bars show the range of increments observed.

We generated pools of size 20, with the $k_N$=10 noise documents and the top $k_G$=10 retrieved by Google for each topic. Then, we added documents returned by the pooling systems to a minimum pool size of 25, 30, 35, and so on, up to the final pools of at least 100 documents. This gives us 17 different pools, each of which can be used to evaluate with the corresponding set of relevance judgments (*trel*, for topic relevance set) from Section 4. We evaluated the 15 student systems for each pool. Then, for each increment of 5 documents in the pool, we calculated the difference in effectiveness for each system between the two pools. The difference is measured as the percentage increased in effectiveness from the smaller to the larger pool, so that it is directly comparable with Zobel's findings (differences between 0.5% and 3.5%, with some observations of up to 19% in TREC-3). Figure 4 shows how the average difference in effectiveness diminishes as the pool size increases.

In the case of NDCG@100, pool sizes larger than 55 show subsequent increments below 1% except in a couple cases (see Table 5). In fact, pools with over 55 documents most increments are below 0.5%. In the case of AP@100 the pool size needs to be at least 85 for the average difference to drop below 1%, while some differences around 2% can still be found. These results show that the pools seem again to be reliable compared to TREC's, although a larger pool size would further increase the reliability of AP@100, P@10 and RR. The patterns observed in EIREX 2010 are extremely similar to the ones found this year: NDCG@100 converges to about 0.5% differences on average, while AP@100 converges to about 1%. P@10 and RR vary between 0.5% and 1.5%. Again, NDCG@100 is shown to be the most stable measure of the four, and therefore the best one to rank systems for grading.

| Pool size | NDCG@100 | | AP@100 | | P@10 | | RR | |
|---|---|---|---|---|---|---|---|---|
| | Mean | Max | Mean | Max | Mean | Max | Mean | Max |
| 20 → 25 | 6.97% | 9.13% | 12.05% | 17.88% | 23.84% | 50% | 18.57% | 32.88% |
| 25 → 30 | 3.15% | 4.74% | 5.5% | 8.83% | 6.67% | 11.59% | 6.67% | 12.25% |
| 30 → 35 | 2.61% | 3.93% | 4.43% | 6.89% | 5.89% | 11.13% | 2.84% | 4.75% |
| 35 → 40 | 2.72% | 7.63% | 3.9% | 13.86% | 5.39% | 21.38% | 5.77% | 26.08% |
| 40 → 45 | 2.19% | 5.73% | 3.14% | 5.08% | 5.1% | 11.59% | 1.79% | 4.31% |
| 45 → 50 | 1.18% | 1.87% | 2% | 2.97% | 3.86% | 7.89% | 2.58% | 6.81% |
| 50 → 55 | 1.35% | 4.73% | 2.11% | 6.4% | 2.88% | 6.03% | 1.26% | 2.73% |
| 55 → 60 | 0.66% | 0.99% | 0.97% | 1.34% | 1.18% | 1.85% | 0.63% | 1.45% |
| 60 → 65 | 0.58% | 1.04% | 1.16% | 1.64% | 1.72% | 2.38% | 1.24% | 3.04% |
| 65 → 70 | 0.45% | 1.03% | 0.99% | 1.22% | 0.92% | 1.57% | 0.65% | 2.29% |
| 70 → 75 | 0.57% | 0.9% | 1.08% | 1.49% | 1.06% | 2.48% | 1.45% | 2.42% |
| 75 → 80 | 0.49% | 0.79% | 1.26% | 1.79% | 1.48% | 2.3% | 0.5% | 3.8% |
| 80 → 85 | 0.49% | 0.89% | 1.05% | 1.71% | 1.56% | 2.41% | 0.88% | 1.93% |
| 85 → 90 | 0.27% | 1.3% | 0.64% | 1.96% | 1.16% | 2.61% | 0.72% | 2.14% |
| 90 → 95 | 0.5% | 1.6% | 0.91% | 2.22% | 0.98% | 1.53% | 1.39% | 5.65% |
| 95 → 100 | 0.26% | 0.67% | 0.67% | 1.29% | 0.64% | 2.06% | 0.26% | 1.52% |

Table 5. Mean and maximum increments observed in NDCG@100, AP@100, P@10 and RR, over all 15 systems, as a function of pool size.

# 6. Conclusions

In 2010 we run the first Information Retrieval Education through EXperimentation (EIREX) experiment to bring TREC-like evaluations to the IR undergraduate classroom [Urbano et al., 2011b]. After the successful experience, this year we proceeded with the second edition of the series: EIREX 2011. With this initiative we get students involved in the whole process of building a search engine and a test collection to evaluate it. Our goal is to introduce students in this kind of laboratory experiments in Computer Science, with a special focus on how to evaluate their systems and analyze the results.

We have described how to adapt TREC's ad-hoc methodology to build such collections for an IR course. The first main difference is that the documents in the collection are gathered after selecting the topics, and not the other way around as usual. The second main difference is related to the pools of documents to judge: the systems developed by the students cannot contribute directly to the pools to prevent cheating, and the judging effort is limited because the students cannot be asked to judge as many documents as we would want. Due to this limitation, the pools are formed differently, with the help of freely available IR tools. The question is whether such small-scale experiments are reliable or not, which is again an excellent question to investigate with the students, so they learn how to analyze them from a critical point of view to look into possible threats to validity [Voorhees, 2002][Urbano, 2011]. The main threats to validity in our case are the inconsistency and incompleteness of relevance judgments. With the EIREX 2010 experiment we found high agreement scores between students, and very high correlations between system rankings when using different sets of relevance judgments; in terms of incompleteness, we estimated that pools of size 100 and different depths were quite reliable and did not seem to affect the evaluation significantly [Urbano et al., 2011b]. This year we also analyzed the effect of incompleteness, and found very similar results. We therefore conclude that the judgments made by students can be trusted, and that the pooling method proposed seems to work reasonably well for these small-scale evaluations.

In 2012 we will probably have many more students, so it is in our plans to build an even larger collection with many more topics. However, students might have problems indexing a much larger collection, so we will investigate a different approach to create the pools. So far, we made relevance judgments on the documents retrieved by the pooling systems for the *complete* collection, but we will make judgments on the documents retrieved for the *biased* collection instead. We found that even the noise documents were relevant in some cases, so there is no reason to believe that documents retrieved for one topic are not relevant for any another topic. The fact that student systems only found about 50% of the documents retrieved for each topic suggests that many of the documents they retrieve are not even judged because they pertain to a different topic. Therefore, we plan on building the biased collection in the same way, but form the judging pools with the documents retrieved after re-indexing only the biased collection. We will try pooling beyond 100 documents per topic and share the judging effort between two students, so we can further reduce the level of incompleteness.